\documentclass[a4paper]{jpconf}
\usepackage{graphicx}

\begin{document}
\title{Numerical cosmology on the GPU with Enzo and Ramses}

\author{C. Gheller, P. Wang, F. Vazza, R. Teyssier}

\address{ETHZ-CSCS Via Trevano 131, Lugano, Switzerland}
\address{NVIDIA, Santa Clara, 95050, US}
\address{Hamburger Sternwarte, Gojenbergsweg 112, 21029 Hamburg, Germany}
\address{University of Zurich, Raemistrasse 71, CH-8006 Zurich}

\ead{cgheller@cscs.ch penwang@nvidia.com franco.vazza@hs.uni-hamburg.de romain.teyssier@gmail.com}

\begin{abstract}

A number of scientific numerical codes can currently exploit GPUs with remarkable 
performance. In astrophysics, Enzo and Ramses are prime examples of such applications. 
The two codes have been ported to GPUs adopting different strategies 
and programming models, Enzo adopting CUDA and Ramses using OpenACC. 
We describe here the different solutions used for the GPU implementation of both cases.
Performance benchmarks will be presented for Ramses. The results of  the usage of the more
mature GPU version of Enzo, adopted for a scientific project within the 
CHRONOS programme, will be summarised. 

\end{abstract}

\section{Introduction}
\label{sec:intro}

The availability of general purpose graphic accelerators (GPUs) has made possible their effective usage for 
scientific computing. Ultimate High Performance Computing (HPC) systems are increasingly 
equipped with GPUs employed not just as graphic accelerators but mainly as computational 
co-processors providing, on suitable classes of algorithms, outstanding performance with 
power efficiency significantly higher than standard CPUs. Supercomputers are thus 
increasingly populated with hundreds or thousands of accelerators that can overlap their computing 
capability with that of CPUs, minimising considerably overall times-to-solution of high-end 
scientific problems. An efficient exploitation of such hybrid architectures 
is not a trivial task, often requiring a full re-design and refactoring
of the code in order to expose its massively parallel algorithmic components,
that can be effectively accelerated by the GPU, and to minimize
or hide any latency related to data transfer between the host and the device.

A number of codes for scientific numerical applications have been recently 
enabled to harness the power of GPUs. In astrophysics the list comprises (but is not limited to) 
the Bonsai, \cite{DBLP:dblp_journals/corr/abs-1204-2280}, Cholla \cite{2014arXiv1410.4194S} and
Gamer \cite{2010ApJS..186..457S} codes. In this paper, we will focus on the Enzo \cite{2014ApJS..211...19B}
and the Ramses \cite{2002A&A...385..337T} codes, which are 
extensively adopted by the computational astrophysics community. Enzo and Ramses are both 
Adaptive Mesh Refinement multi-species codes, designed to describe the evolution of 
cosmological structures, galaxies, gas clouds etc. The two codes can run on massive parallel
HPC systems thanks to an effective MPI based implementation. 
An overview of the main features of Enzo and Ramses will be given in 
section \ref{sec:overview}.

Enzo and Ramses have been ported to GPUs adopting different strategies and programming models, 
Enzo exploiting CUDA and Ramses using OpenACC. The different solutions used for the GPU refactoring will be described 
in the following sections \ref{sec:ramses} and \ref{sec:enzo}. 
Note that a CUDA based implementations of the radiative transfer module of Ramses has
been developed by \cite{2010ApJ...724..244A}.

In section \ref{sec:ramses}
we will focus on the re-design and refactoring work performed on Ramses, for 
which a number of benchmarks comparing the GPU and the CPU versions of the codes will be presented, 
highlighting the enhancements obtained with the refactoring and pointing out the 
main factors currently limiting the performance improvements.

The more mature Enzo's GPU implementation will be introduced in section \ref{sec:enzo}.
It is currently used in a project within the 
CHRONOS\footnote[1]{http://www.cscs.ch/user\_lab/allocation\_schemes/chronos\_projects/index.html} programme, 
supporting high-end computational physics applications. 
The project aims at studying the origin and evolution of the magnetic field within large-scale structures 
and exploits the MHD CUDA enabled solver, allowing to overcome the bounds related to 
memory size, while maintaining the computational time reasonable. An overview of the achieved performance
will be presented together with some notable scientific outcomes. 

Both Ramses benchmarks and Enzo runs have been performed on the Piz Daint HPC system at ETHZ-CSCS,
a Cray XC30 supercomputer accounting for more than 5000 computing nodes, each equipped with an 8-core 
Intel SandyBridge CPU (Intel Xeon E5-2670) and an NVIDIA Tesla K20X GPU.

Section \ref{sec:conclusions} will summarize the work and draw the main conclusions.

\section{Enzo and Ramses Overview}
\label{sec:overview}

Enzo and Ramses are both Adaptive Mesh Refinement (AMR) codes designed to solve 
a broad variety of astrophysical problems. 
The AMR approach is the key feature for the two codes, also in terms of GPU enabling strategy. 
AMR provides high spatial resolution 
only where this is actually required, thus ensuring minimal memory usage and computational
needs. The codes adopt different AMR solutions, namely Structured AMR (SAMR) by Enzo 
and Fully Threaded Tree (FTT \cite{Khokhlov1998519}) by Ramses. 
With the SAMR approach, the computational domain is represented by a
rectangular grid on which a solution of the underlying problem is computed.
Regions requiring additional resolution are identified by some suitable criteria
and covered by a disjoint union of rectangles (patches), which are then refined by integer factors
(usually a factor of two is adopted). 
The solution is then computed on the superposition of patches. This process is used recursively 
and refinement are applied in space as well as in time. 
Similarly, FTT supports local refinements of the spatial domain, but with an 
unstructured approach, dynamically creating refinements on a cell-by-cell basis, 
the mesh being able to conform to complex boundaries.   
Spatial and time refinement criteria are the same as for SAMR.

The AMR hierarchy management is organized as a tree. In the case of Enzo, each node of the
tree represents a patch  which is pointed by the coarser father patch, and points to 
the higher resolution nested son patches. In the case of Ramses, each cell stores information
about the parent (at lower resolution), neighbouring (at the same resolution) and children 
(at higher resolution) cells. In this way, the tree can be quickly traversed during 
various search operations. Comparing the two approaches, FTT optimizes memory usage and
can provide efficient parallel implementations. However, memory access is not as efficient 
and simple as for the regular patches supported in SAMR. 

On the AMR mesh, Enzo and Ramses solve the equations describing the dynamics of the two main matter component in the 
universe, dark and baryonic matter (DM and BM), driven by the gravitational 
field, generated by the combined mass distributions.  
The two codes use a particle-mesh N-body method to follow the dynamics of the collisionless DM component. 
The BM component is represented as an ideal fluid, discretised on the AMR mesh.
Its behaviour is described solving the Euler equations through different numerical
approaches (e.g. the Piecewise Linear Method, PLM \cite{1985JCoPh..59..264C}, the Piecewise Parabolic Method, PPM \cite{cw84} or the TVD \cite{Yee1985327} schema).
The gravitational potential is calculated solving the Poisson equation
through a multigrid or a combined multigrid plus fast fourier
transform approach.
A variety of further physical processes is included in the two codes. Standard
hydrodynamics is extended to magneto-hydrodynamics (MHD). Sub-grid phenomena,
like star formation or supernovae explosions, are modelled. Additional basic physics,
like cooling processes or radiative transfer, are supported.

The GPU enabling of the two codes follows distinct strategies,
driven by the AMR approach, and programming models, as described in the 
following two sections. 

\section{Ramses OpenACC Implementation and Benchmarks}
\label{sec:ramses}

For Ramses, an incremental GPU porting strategy has been defined,
extending and optimizing progressively the fraction of the code ported on the accelerator.
The hydrodynamics solver has been identified as a primary target for Ramses' GPU implementation,
being among the most computational demanding code's components and, at the same time, 
solving a local problem, ideally fitting the accelerator’s architecture. This first step is being followed
by the progressive porting of all the other algorithmic components.
This approach ideally fits the OpenACC programming model, 
which, furthermore, fully supports Fortran 90
and reduces the impact on the source code, minimizing the effort for 
its management and maintenance. OpenACC has been adopted as the solution
for Ramses GPU refactoring.

\begin{figure*}
\begin{center}
\includegraphics[width=1.0\textwidth]{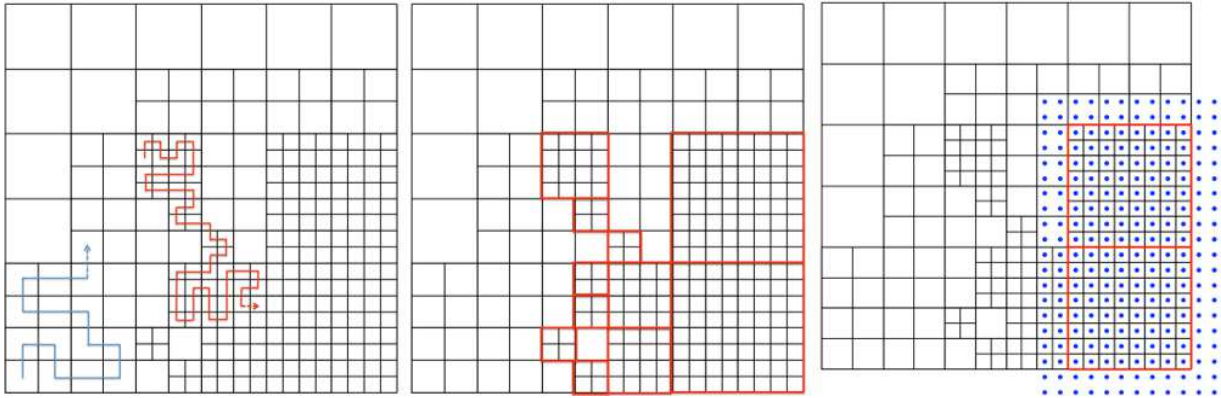}%{images/ramses-amr.eps}
\caption{
Ramses' hydro solver implementation on the GPU: step 1 (left), cells are ordered according to 
a space filling curve at each level; step 2 (centre), regular rectangular patches are created; 
step 3 (right), patches are gathered composing large chunks and hydro equations are solved.  
}
\label{fig:ramses_amr}
\end{center}
\end{figure*}

The main challenge for an effective design of the hydro GPU algorithm is represented by the FTT 
based data management. The FTT approach
allows a prompt identification of sibling and parent cells. However data results 
scattered in memory and its access, even in the case of neighbouring cells,
is often on non contiguous memory locations. This represents a major issue  in terms of performance
for any processor relying on a cache based memory hierarchy. In order to 
circumvent this problem, Ramses gathers in a buffer all the data needed for 
the update of a few cells at a time. The resulting buffer's size is $N_v\times 6\times6\times6$ cells, $N_v$ being the number of cells to update 
set to $N_v \sim 10$, in order to 
fit the cache memory hierarchy. This solution is not suitable to 
the GPU, since larger contiguous data chunks are necessary to maximise the GPU occupancy, minimizing 
at the same time the amount of memory accesses. A simple increase of $N_v$ would not be effective,
preserving the number of accesses to main memory. Furthermore, data would be replicated 
unnecessarily, due to the $6^3$ cubes building process. 
Hence, for the GPU a specific procedure has been designed and can be summarised as follows (see also
Figure \ref{fig:ramses_amr}):
\begin{itemize}
\item at a given level of refinement, AMR cells are ordered according to a space filling curve 
based layout;
\item cells are grouped in compact (i.e. with no holes) cubic patches of maximum possible size;
\item neighbouring patches are grouped in large chunks collecting all the data necessary for the hydro equations integration,
(i.e. inner cells plus boundary cells, properly calculated through interpolation if the patch
adjoins cells at a different refinement level;
\item hydro equations are solved on the chunk of data
\item next chunk of data at the same level starts to be created or, if the level has been solved,
calculation on the next coarser level begins.
\end{itemize}

Data transfers to and from the GPU have been minimised by transferring only necessary data 
i.e. data belonging the refinement level being updated) are moved to and from the GPU. 

\begin{table}[h]
\caption{Comparison of CPU and GPU performance on a representative cosmological run with a $256^3$
base mesh and 8 levels of refinement. All the timings are in seconds.}
\begin{center}
\begin{tabular}{lllll}
\br
Code version & Hydro (tot) & Hydro (calc) & Hydro (copy) & Total\\
\mr
CPU (1 core)   & 56218 & 56218 & NA  & 155662 \\
CPU (16 cores) & 2918  &  2918 & NA  & 8775   \\
1 GPU          & 3009  &  2270 & 739 & 104811 \\
16 GPUs        & 179   &  115  & 64  & 5718 \\
\br
\end{tabular}
\label{tab:bench}
\end{center}
\end{table}

In Table \ref{tab:bench} we present some benchmarks performed to compare the CPU and OpenACC 
versions of the code in a representative cosmological simulation, with a $256^3$ base grid and 8 levels of
refinement. The tests were run on a single CPU core, on a full CPU socket (two CPUs
sharing a common memory, for a total 
number of 16 cores), on a single GPU and on 16 GPUs (one GPU per computing node). Column 2 of the Table, compares 
the timings to run the hydro kernel. On overall, the GPU performs as 16 cores and about 18 times faster than a single core.
A fraction of about 75\% of the time on the GPU is spent on computation (column 3), 
while the remaining time is spent on data transfer. Due to the characteristics of the algorithm,
this overhead cannot be hidden exploiting asynchronous
processes, overlapping computation and data transfer. It will be reduced only porting a larger
fraction of the code on the GPU, thus reducing the need for data transfers.  
The impact of the GPU on the total run time (column 5) is, of course, limited by 
the fact that only the hydro kernel is running on the accelerator, stressing, once more, 
the importance of having the majority of the code ported on the GPU. 
Finally, by comparing the results for 1 and 16 GPUs, we see how, for the case under investigation,
the good scalability of the hydro kernel is preserved, despite the work being 
distributed among 16 distinct nodes, communicating through the network. 
Again, the copy time is a major source of overhead, the
data to be transferred to the device increasing with the number of GPUs 
relatively to the amount of computation per device. 

\section{Enzo MHD GPU Simulations}
\label{sec:enzo}

Enzo's MHD solver, as well as the PLM/PPM hydro solver, have been ported
to NVIDIA's CUDA framework (Wang et al. 2010; The Enzo Collaboration et al. 2013). Data
is off-loaded to the GPU asynchronously on a patch-by-batch basis, effectively overlapping computation
and data transfer. All fluxes necessary to update the patch's cells are computed on the GPU and
only fluxes required for flux correction necessary for multilevel, adaptive time-step, calculations
are moved back to the CPU, representing the only additional data transfer overhead.
The porting onto GPUs replaced many shared temporary arrays of the CPU version into
larger temporary arrays that are not shared among loop iterations,
and exposed the massive parallelism of the algorithm. The larger memory usage
is not a major issue, since updated patches are immediately transferred back to the CPU,
freeing the corresponding GPU memory.
For further details on the porting onto CUDA, we refer the reader to \cite{wang10} and
\cite{2014ApJS..211...19B}.

\begin{figure*}
\begin{center}
\includegraphics[width=0.95\textwidth]{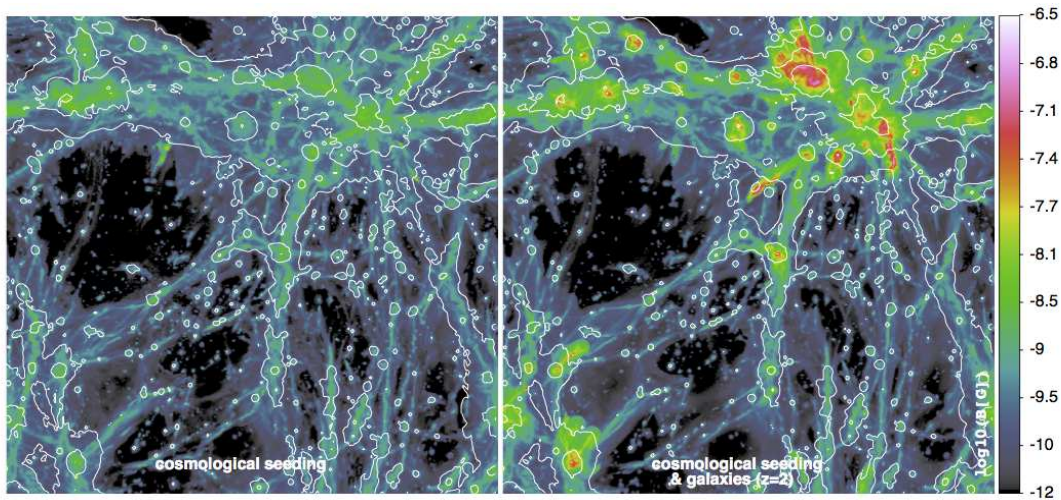}%{images/B-proj.eps}
\caption{Projected maps of mass-weighted magnetic field intensity (with overlaid logarithmic contours of projected temperature) for a $1200^3$ simulations of a (25 Mpc)$^3$ volume at $z=0$, with a cosmological weak magnetic field initialised at $z=30$ (left) or with the additional
release of magnetic loops from ``galaxies" in the volume at $z=2$ (right). }
\label{fig:franco_map}
\end{center}
\end{figure*}

Enzo's GPU implementation is currently used in a CHRONOS project,
aiming at studying the origin of cosmic magnetism in general
and the observed  distribution of magnetic fields in large scale structures.
This is in fact an astrophysical puzzle.  On one hand observations provide evidence for magnetic field strengths of up to a few $\sim \rm \mu G$ in galaxy clusters and galaxies (e.g. \cite{fe08}), on the other the origin of such strong fields is unclear, given the low upper limits on the primordial magnetic field at the epoch of the Cosmic Microwave Background, and  the uncertainty on fields generated by primordial mechanisms is  still huge  (e.g. \cite{wi11}).\\
Cosmological simulations already proved that  significant field amplification in galaxy clusters can be achieved by starting from weak cosmological fields and following turbulent small-scale dynamo (e.g. \cite{do08} and references therein). However, field amplification in the outer part of clusters and in filaments of the cosmic web is still poorly investigated. 
Recently, in  \cite{va14mhd}, we tackled this problem. We uses the GPU-ported version of Enzo-MHD  to achieve a high-resolution view of magnetic field amplification in the cosmic web. As we were mostly interested in the outer region of clusters
and filaments, we primarily used large unigrid simulations, which are less affected by numerical dissipation compared to AMR runs and allow complete surveys of large cosmic volumes. \\
Running on Piz-Daint on 512 to 2048 computing nodes using, due to memory constraints, one to four tasks per node, we  investigated various details of the numerical modelling of cosmic magnetism: from dependence on cluster/filament mass to the convergence with spatial resolution. Furthermore, we explored the role of additional seeding by magnetized jets from active galactic nuclei within clusters.
In this project we produced the current largest cosmological MHD simulation to date ($2400^3$ cells and dark matter particles), as well as the most complete resolution study to date on the amplification in a cosmic filaments. In particular, the use of the GPU allowed us to
complete such large unigrid run in $\approx 4.5 \cdot 10^6$ core hours with 512 nodes, cutting by a factor $\sim 4$ the total computational time compared to the CPU version of the code. \\
Figure \ref{fig:franco_map} (left) shows the projected magnetic field strength across a $1200^3$ box simulating (25 Mpc)$^3$, at the final timestep (redshift $z=0$). Magnetic fields are advected and amplified where gas matter concentrates, however their intensity never becomes
dynamically relevant compared to the gas energy. 
For the small simulated galaxy clusters/groups, the maximum measured magnetic field is $\sim 0.1 \mu G$, while the typical strength in filaments is instead $\sim 0.001 \mu G$. The right panel in Figure  \ref{fig:franco_map} shows the potential role played by the additional release of magnetic fields from galaxies inside galaxy clusters. While this can boosts the maximum field inside clusters up to $\sim 1 \mu G$ (i.e. close to observations),  this additional seeding mechanisms does not increase significantly the magnetisation of filaments.

These results suggest that the memory of the seeding event(s) must be kept within filaments, as
their evolution is dominated by compression and not by strong turbulence (which would erase any information on the initial seed field, as in galaxy clusters). We are presently working together with radio astronomers involved in the design and testing of the Square Kilometer Array, to assess which crucial
information of cosmic magnetic fields will be accessible with the future configurations planned for the SKA, 
which might be able to give us crucial information of the initial seeding mechanisms of large-scale magnetic fields in the Universe \cite{brown11}.

\section{Conclusions}
\label{sec:conclusions}

This paper has summarised the GPU implementation of the Enzo and Ramses codes. Although only part 
of both codes is currently implemented to exploit GPU empowered supercomputers, they have already 
proved the effectiveness of solutions capable of running on hybrid accelerated systems. Enzo has been used
to run a number of large MHD simulations overcoming the limitations posed by memory constraints
of current CPUs keeping, thanks to the GPU, the computational time acceptable. Ramses, despite
the good results obtained for the hydro kernel,  
is pointing out how the porting of the whole application is often necessary for the full 
exploitation of the accelerator, minimizing the host-device data transfer overheads. On-going work is finalized 
to enable the cooling, MHD and gravity modules to the GPU.

\section*{References}
%\bibliography{cla-ref}
\bibliography{paper}

\providecommand{\newblock}{}
\begin{thebibliography}{10}
\expandafter\ifx\csname url\endcsname\relax
  \def\url#1{{\tt #1}}\fi
\expandafter\ifx\csname urlprefix\endcsname\relax\def\urlprefix{URL }\fi
\providecommand{\eprint}[2][]{\url{#2}}

\bibitem{DBLP:dblp_journals/corr/abs-1204-2280}
B\'{e}dorf J, Gaburov E and Zwart S~P 2012

\bibitem{2014arXiv1410.4194S}
{Schneider} E~E and {Robertson} B~E 2014 {\em ArXiv e-prints\/}
  (\textit{Preprint} \eprint{1410.4194})

\bibitem{2010ApJS..186..457S}
{Schive} H~Y, {Tsai} Y~C and {Chiueh} T 2010 {\em \apjs\/} {\bf 186} 457--484
  (\textit{Preprint} \eprint{0907.3390})

\bibitem{2014ApJS..211...19B}
{Enzo Collaboration} 2014 {\em \apjs\/} {\bf 211} 19 (\textit{Preprint}
  \eprint{1307.2265})

\bibitem{2002A&A...385..337T}
{Teyssier} R 2002 {\em \aap\/} {\bf 385} 337--364 (\textit{Preprint}
  \eprint{astro-ph/0111367})

\bibitem{2010ApJ...724..244A}
{Aubert} D and {Teyssier} R 2010 {\em \apj\/} {\bf 724} 244--266
  (\textit{Preprint} \eprint{1004.2503})

\bibitem{Khokhlov1998519}
Khokhlov A 1998 {\em Journal of Computational Physics\/} {\bf 143} 519 -- 543
  ISSN 0021-9991

\bibitem{1985JCoPh..59..264C}
{Colella} P and {Glaz} H~M 1985 {\em Journal of Computational Physics\/} {\bf
  59} 264--289

\bibitem{cw84}
{Colella} P and {Woodward} P~R 1984 {\em Journal of Computational Physics\/}
  {\bf 54} 174--201

\bibitem{Yee1985327}
Yee H, Warming R and Harten A 1985 {\em Journal of Computational Physics\/}
  {\bf 57} 327 -- 360 ISSN 0021-9991
  \urlprefix\url{http://www.sciencedirect.com/science/article/pii/0021999185901834}

\bibitem{wang10}
{Wang} P, {Abel} T and {Kaehler} R 2010 {\em \na\/} {\bf 15} 581--589
  (\textit{Preprint} \eprint{0910.5547})

\bibitem{fe08}
{Ferrari} C, {Govoni} F, {Schindler} S, {Bykov} A~M and {Rephaeli} Y 2008 {\em
  \ssr\/} {\bf 134} 93--118 (\textit{Preprint} \eprint{0801.0985})

\bibitem{wi11}
{Widrow} L~M, {Ryu} D, {Schleicher} D, {Subramanian} K, {Tsagas} C~G and
  {Treumann} R~A 2011 {\em ArXiv e-prints\/} (\textit{Preprint}
  \eprint{1109.4052})

\bibitem{do08}
{Dolag} K, {Bykov} A~M and {Diaferio} A 2008 {\em \ssr\/} {\bf 134} 311--335
  (\textit{Preprint} \eprint{0801.1048})

\bibitem{va14mhd}
{Vazza} F, {Br{\"u}ggen} M, {Gheller} C and {Wang} P 2014 {\em \mnras\/} {\bf
  445} 3706--3722 (\textit{Preprint} \eprint{1409.2640})

\bibitem{brown11}
{Brown} S, {Emerick} A, {Rudnick} L and {Brunetti} G 2011 {\em \apjl\/} {\bf
  740} L28+ (\textit{Preprint} \eprint{1109.0316})

\end{thebibliography}

\end{document}